\documentclass[12pt]{iopart}

\usepackage{graphicx}

\begin{document}

\title{The effect of Coulomb interaction at ferromagnetic-paramagnetic
  metallic perovskite junctions.}

\author{R. Allub$^{1}$\thanks{ Member of the Carrera del Investigador
    Cient\'{\i}fico del Consejo Nacional de Investigaciones
    Cient\'{\i}ficas y t\'{e}cnicas (CONICET).}, J. D. Fuhr$^{1}$%
  , \ M. Avignon$^{2}$, and B. Alascio$^{1}$ }

\address{$ ^{1} $ Centro At\'{o}mico Bariloche, (8400) S. C. de
  Bariloche, Argentina.}

\address{$ ^{2} $ Institut N\'eel, CNRS and Universit\'e Joseph
  Fourier, BP 166, 38042 Grenoble Cedex 9, France.}

\begin{abstract}
  We study the effect of Coulomb interactions in transition metal
  oxides junctions. In this paper we analyze charge transfer at the
  interface of a three layer ferromagnetic-paramagnetic-ferromagnetic
  metallic oxide system.  We choose a charge model considering a few
  atomic planes within each layer and obtain results for the magnetic
  coupling between the ferromagnetic layers. For large number of
  planes in the paramagnetic spacer we find that the coupling
  oscillates with the same period as in RKKY but the amplitude is
  sensitive to the Coulomb energy. At small spacer thickness however,
  large differences may appear as function of : the number of
  electrons per atom in the ferromagnetics and paramagnetics
  materials, the dielectric constant at each component, and the charge
  defects at the interface plane emphasizing the effects of charge
  transfer.
\end{abstract}


\submitto{\JPCM}



\section{INTRODUCTION}

Magnetic multilayer films have attracted the attention of the physics
community since the discovery of "Giant magnetoresistance " (GMR) in
metallic superlattices by Baibich et al. \cite{ba} and have been the
subject of intensive basic and applied research. The fundamental
physics giving rise to GMR has been clarified, the role of RKKY
interactions being prominent.  Application to read heads of magnetic
memories is already a reality. The oscillations in the magnetic
coupling have been beautifully derived by analogy to the de Haas Van
Alfven effect by D. M. Edwards et al.\cite{Ed} and were used to
calculate realistically the magnetic coupling between Co layers in
CoCuCo trilayers \cite{Ma}. A comprehensive discussion of the problem
of interlayer coupling is given in Ref. \cite{Bruno}. Studies of
metal to insulating oxides interfaces based on Density Functional Theory (DFT) are
reported in Ref. \cite{Oley}

On the other side, the discovery of "Colossal magnetoresistance" by
von Helmholt et al \cite{helm} in La$_{1-x}$Sr$_{x}$MnO$_{3}$ type
compounds, has also polarized research on these materials, their phase
diagrams as a function of composition and temperature have been
determined, but even though there has been considerable progress, the
basic physics of the bulk materials is not yet quite completely
understood \cite{Pe}. Furthermore, several magnetic oxide
heterostructures based on colossal magnetoresistance materials have
been the subject of important research, including magnetic coupling
mediated by a metallic spacer \cite{Ni},\cite{Mara}. We study here the
effect of the charge transfer (CT) at the interface on the magnetic
coupling. RKKY being a perturbation theory can not be applied to these
materials where the coupling energy between moments and conduction
electrons is larger than the bandwidth. The theories mentioned before
are appropriate for metallic magnets were screening is strong and
Coulomb interactions play a minor role in the charge transfer at the
interface. In this paper we focus on a trilayer formed by
ferromagnetic-paramagnetic ferromagnetic ionic compounds and alloys
where the Coulomb interaction dominates the charge transfer at the
interface. We calculate self-consistely the charge profile in terms of
a minimum set of parameters and study the effect of these parameters
on the magnetic coupling between ferromagnetic layers. To this end, we
build a Hamiltonian that contains the essential features of a
fully-polarized ferromagnet (FF) and a paramagnetic metallic (PM)
spacer and parametrize the interfaces in the simplest possible way.

\section{MODEL}

We study a model for electrons moving in a three layer structure
(FF-PM-FF) formed by two N-cell units (N-c.u.) of half-metallic
ferromagnetic perovskites, as for example La$_{1-x}$Sr$_{x}$MnO$_{3}$
($0.2\leq x\leq 0.5$) separated by M-c.u. of non-magnetic metal, as
LaNiO$_{3}$. The Mn and Ni sites form a simple cubic structure and we
take the interfaces perpendicular to a cubic axis. La (Sr) at the
center of the cubes form an other interpenetrating cubic lattice with
opposite charge, so that La(Sr) planes are in between the Mn
planes. We will consider this alternating planar structure as shown
schematically in Fig. 1.
\begin{figure}[th]
\begin{center}
\includegraphics[width=7cm]{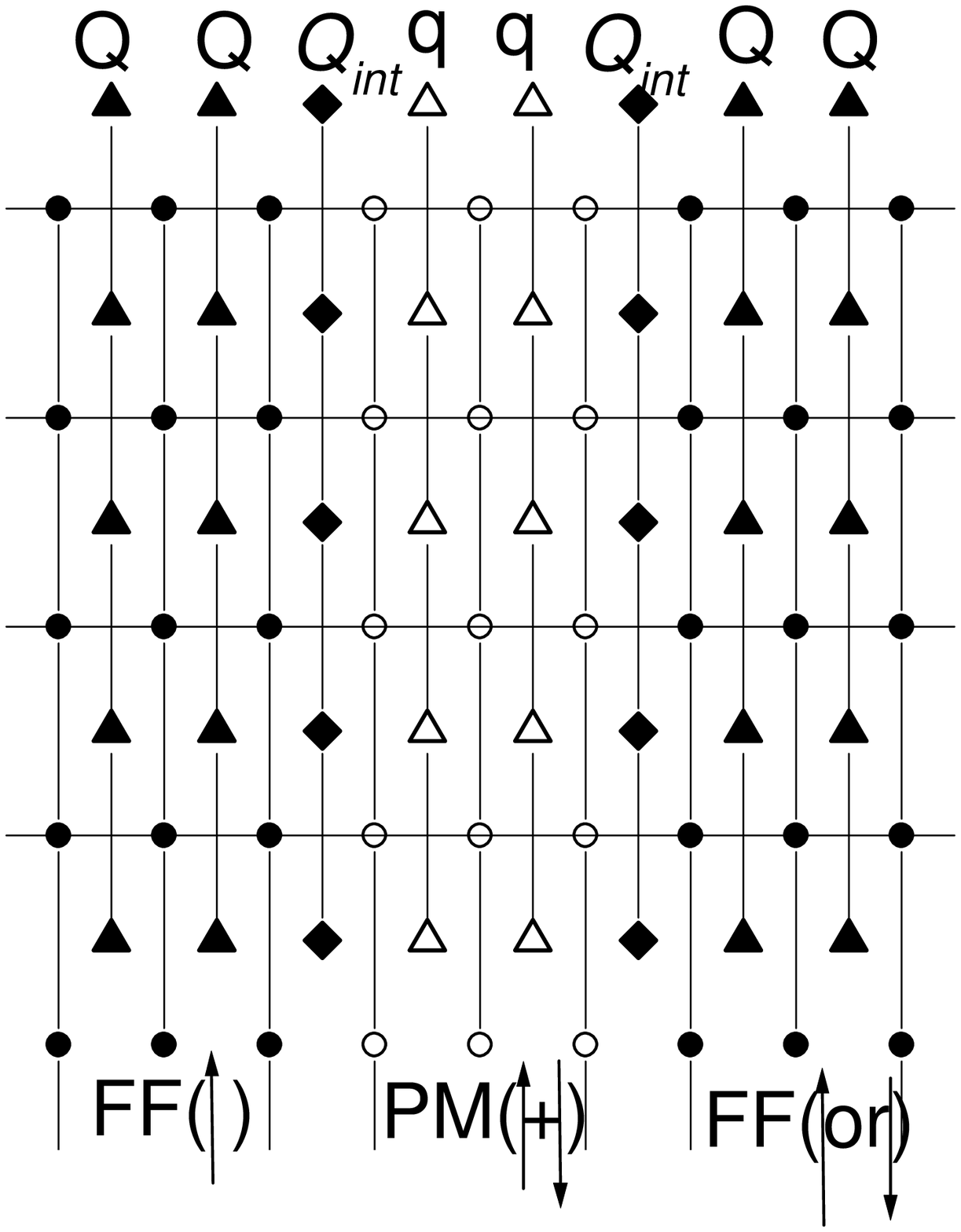}
\end{center}
\caption{Schematic figure of the model used in this study. Solid and
  open circles show the positions of electron lattice sites for FF and
  PM materials respectively. Solid and open triangles show the
  positions of positive lattice sites for FF and PM
  respectively. Solid diamond show the positions of positive charge
  $Q_{int}$ in both planes adjacent to the PM region.}
\label{fig1}
\end{figure}
To describe the charge transfer between the different layers we follow
the pioneer papers of Gorkov and Kresin \cite{GK} and Okamoto and
Millis \cite{OM}, and used more recently by Brey \cite{BR} to study
manganite-insulator oxide interfaces. Since charge transfer at the
interfaces is determined by the competition between kinetic energy and
Coulomb energy we define a Hamiltonian for the kinetic energy of
manganites $H_{N}$, and for the spacer $H_{M}$ and a Coulomb
Hamiltonian $H_{Coul}$ that accounts for the Coulomb interactions
between electrons and the positive charges background, and between
themselves. Because of the large intrasite Coulomb interaction and
exchange energies, double occupation of the e$_{g}$ orbitals is
inhibited.  To include this fact in the model Hamiltonian we use a
single spinless orbital in the Hamiltonian following previous models
for manganites Ref.\cite{allub}. In manganites each Mn ion has three
localized $3d$ electrons in the $t_{2g}$ orbitals which, due to Hund's
rule, produce a local spin $S=3/2$, while the additional $e_{g}$
electrons are itinerant and have their spin parallel to the local spin
again due to Hund's rule. When both manganites layers are strongly
ferromagnetic, the conduction electrons can be completely spin
polarized parallel to the magnetization in each manganite region, the
situation we are considering here. The itinerant electrons in the
right and left FF regions can be modelled by a tight-binding
Hamiltonian:
\begin{equation}
H_{N}=\sum_{i,j}\epsilon _{i}c_{i,j}^{\dagger }c_{i,j}-t\sum_{<i,j,\delta
,\delta ^{\prime }>}c_{i,j}^{\dagger }c_{i+\delta ,j+\delta ^{\prime }}
\end{equation}
where $i$ identifies the planes and runs between $1$ and $N$ in the
left FF layer and between $N+M+1$ and $2N+M$ in the right one, $\delta
$ run over nearest neighbors (n.n) layers, $j$ covers each plane in
the layer, and $ \delta ^{\prime }$ run over n.n in the plane $j$.

For the spacer, with LaNiO$_{3}$ in mind, a metallic paramagnet\cite
{LNO}, we take again a similar single orbital Hamiltonian $H_{M}$ with
the only difference that now it includes the spin of the electrons,
and $i$ will run from $N+1$ to $N+M$; all other indices keep the same
meaning.\ For simplicity we take a single value for the hopping
parameter $t$ between all nearest-neighbor orbitals. We then consider
either ferro (F) or antiferromagnetic (AF) alignments between the left
and right FF layers to calculate the difference of energy between the
two configurations. This energy difference arises from the fact that:
in the ferromagnetic alignments electrons with spin up can move freely
within the three layers and those with spin down are confined to the
spacer, while in the case of AF alignments, electrons with spin up
move between the first and second layer and those with spin down
between the second and third.

The electrons move in a background of positive charges centered at the
interpenetrating lattice which will be defined later. The average
number of electrons per site $n$ is fixed such that the whole system
is neutral.  Electrons hopping between all n.n orbitals feel a
potential arising from the extra positive charges and the
electron-electron repulsion.  The diagonal energies $\epsilon _{i}$ in
Eq. (1) will result here from a self-consistent calculation involving
the Coulomb energy due to the charges at the positive and negative
interpenetrating lattices. This is the main contribution of this paper
to the study of magnetic coupling between layers.  The Coulomb
interaction, which is the most relevant ingredient in the model, takes
the following form:
\begin{eqnarray}
  H_{Coul} &=&-\sum_{i,l}\frac{e^{2}n_{i}q_{l}}{\epsilon ~\left\vert
      R_{i}-R_{l}\right\vert }+\frac{1}{2}\sum_{i\neq j}
  \frac{e^{2}n_{i}n_{j}}{
    ~\epsilon \left\vert R_{i}-R_{j}\right\vert }  \nonumber \\
  &&+\frac{1}{2}\sum_{m\neq l}\frac{q_{m}q_{l}e^{2}}{\epsilon ~\left\vert
      R_{m}-R_{l}\right\vert },
\end{eqnarray}
where $i,j$ refer to the electron lattice while $l,m$ refer to the
positive lattice, both indexes run over the whole system. The first
term represents the attraction between electrons and positive charges
while the second and third correspond to the repulsion between charges
of the same sign. $ \epsilon $ is the dielectric constant which can be
quite high. The electronic charge at each FF site or PM site $i$ is
$-en_{i}$, $n_{i}$ will be calculated self consistently by mean field
theory. To simplify the notation we use a single parameter
$U=\frac{e^{2}}{\epsilon \ast a}$ where $ a $ is the lattice
parameter. The factor $eq_{l}$ represents the average charge per unit
cell in the positive lattice. To represent the three layer structure
we take three different values for $q_{l}$: $q_{l}=Q$ in the FF
layers, $q_{l}=q$ for the spacer, and to model possible distortions of
charge or structure at the plane between FF and PM we use as a first
approach $\ q_{l}=Q_{int}$. Here, $Q_{int}$ is a single interface
parameter which allows us to introduce different charge in both
positive planes adjacent to the PM. Note that since positive planes
are intercalated between electron planes the number of positive planes
is $2N+M-1$. In Fig. 1 we show schematically the charge structure in a
N=3, M=3 FF-PM-FF example.

To consider the example of the manganites and nickelate FF-PM-FF
structure as mentioned above, we would define the positive charges in
the following way. The $e_{g}$-orbitals are empty for Mn$^{4+}$ and
occupied with one electron in Mn$^{3+}$. In this way the neutral
Sr$^{2+}$Mn$^{4+}$O$_{3}^{2-}$ background has ionic character with no
conduction electrons and\ we will describe the charges in our system
as the additional charges with respect to this background.\ La has a
nominal valence $3+$ therefore each La produces an excess charge $+1$
together with one conduction electron. So in La$_{1-x}$
Sr$_{x}$MnO$_{3}$ the number of conduction electrons per unit cell is
$1-x$ and $Q=1-x$. We take the same reference background of positive
charges also in LaNiO$_{3}$, so that now in the spacer one would have
one itinerant electron per nickel site and $q=1$.

\section{RESULTS}

Since the model contains several parameters, to fix ideas we present
the results choosing values of the parameters appropriate to describe
the charge effects in the three layer systems investigated in
Ref. 6. Accordingly we chose $N=12$, $Q=2/3$, and $q$ $=1$ in most of
our results. We adjust the Fermi energy so that the total number of
electrons equals the positive charge ($(2N+M)n=$
$2(N-1)Q+(M-1)q+2Q_{int}$). We take $t$ as the unit of energy. \

\begin{figure}[ht]
\begin{center}
\includegraphics[width=7cm]{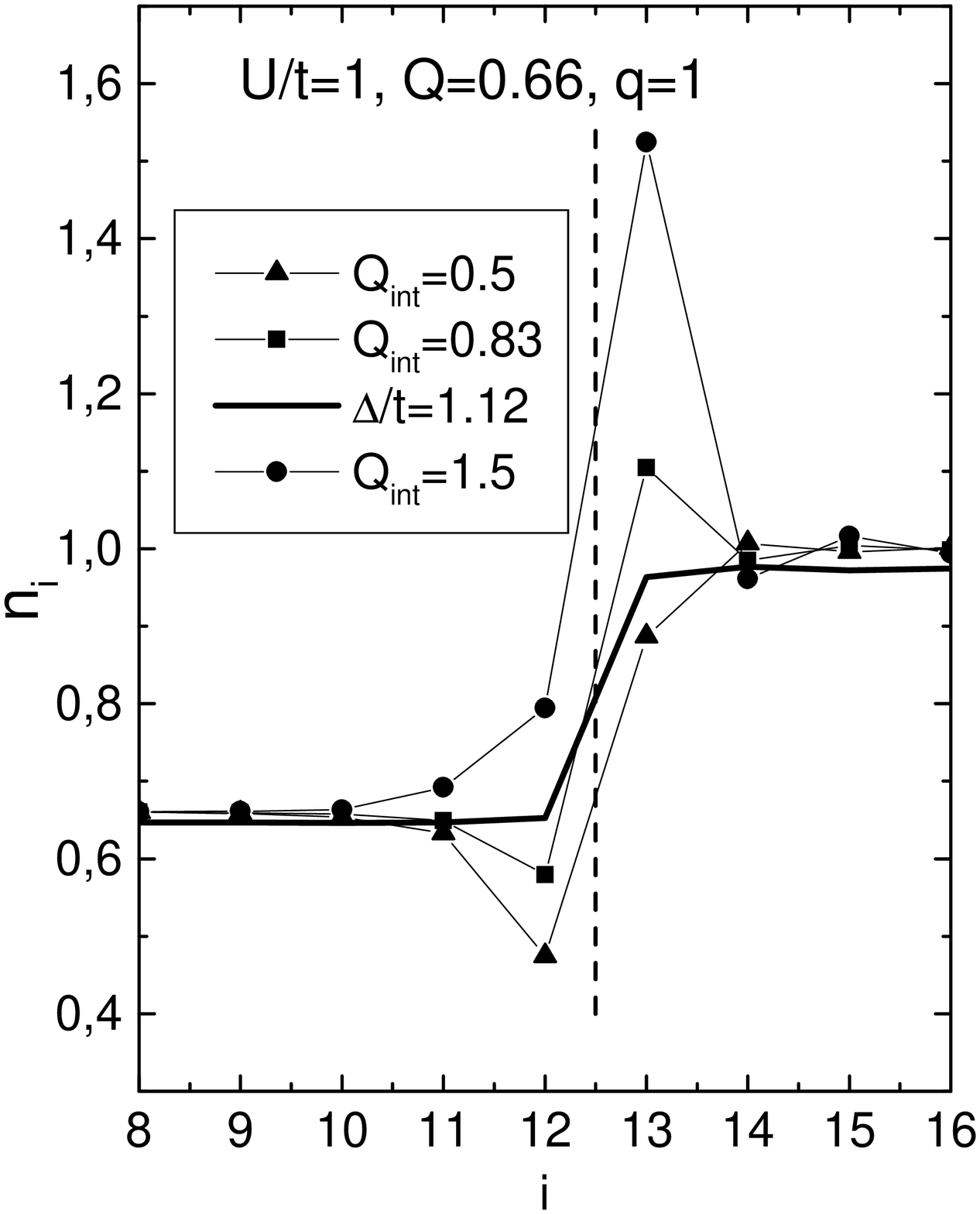}
\end{center}
\caption{The mean value of number of particles per site n$_{i}$ at the
  $i$ -th plane as a function of $i$ for $N$=12, $M$= 14, $U/t$=1, $Q$
  =0.66, $q$ =1, and three different values of $Q_{int}$ in the AF
  alignments. Vertical dash line ($i=$12.5) is a guide to the eye to
  show the interface plane between FF and PM materials. The broad line
  corresponds to a step potential $ \Delta /t=1.12.$}
\label{fig2}
\end{figure}

We begin by looking to the effect of the charge of the interface plane
$ Q_{int}$ on the electron charge distribution. We find that the
charge distribution is almost insensitive to the magnetic alignment of
the FF layers. However, an important charge transfer occurs in the
planes very near to the interface FF / PM. In Fig.~\ref{fig2} we show
$n_{i}$ as a function of $i$ for $M=14$, $U/t=1$, $Q$ $=0.66$, and
three different values of $ Q_{int}$. Not surprisingly the average
electron charge of the nearest planes follows the excess or defect of
charge of the interface plane, however the charge of the first
separator plane exceeds its mean value ($\sim q$) by 10\% at average
$Q_{int}=0.83$ and can go up or down according to $Q_{int}$.  On the
contrary, the charge at the first FF layer is lower than $Q$ while one
would expect that the charge would accumulate on both sides of the
interface. This seemingly surprising result is a consequence of the
fact that each site at the separator can be doubly occupied while the
FF \ sites can be only singly occupied. Since many of the FF are
manganites and they are extremely sensitive to the average charge this
fact could affect strongly the magnetic properties of the
interface. Here we assume that all atomic planes at the FF are fully
polarized. The effects of charge inhomogeneity in the FF planes will
be reported in a different publication.

\begin{figure}[ht]
\begin{center}
\includegraphics[width=7cm]{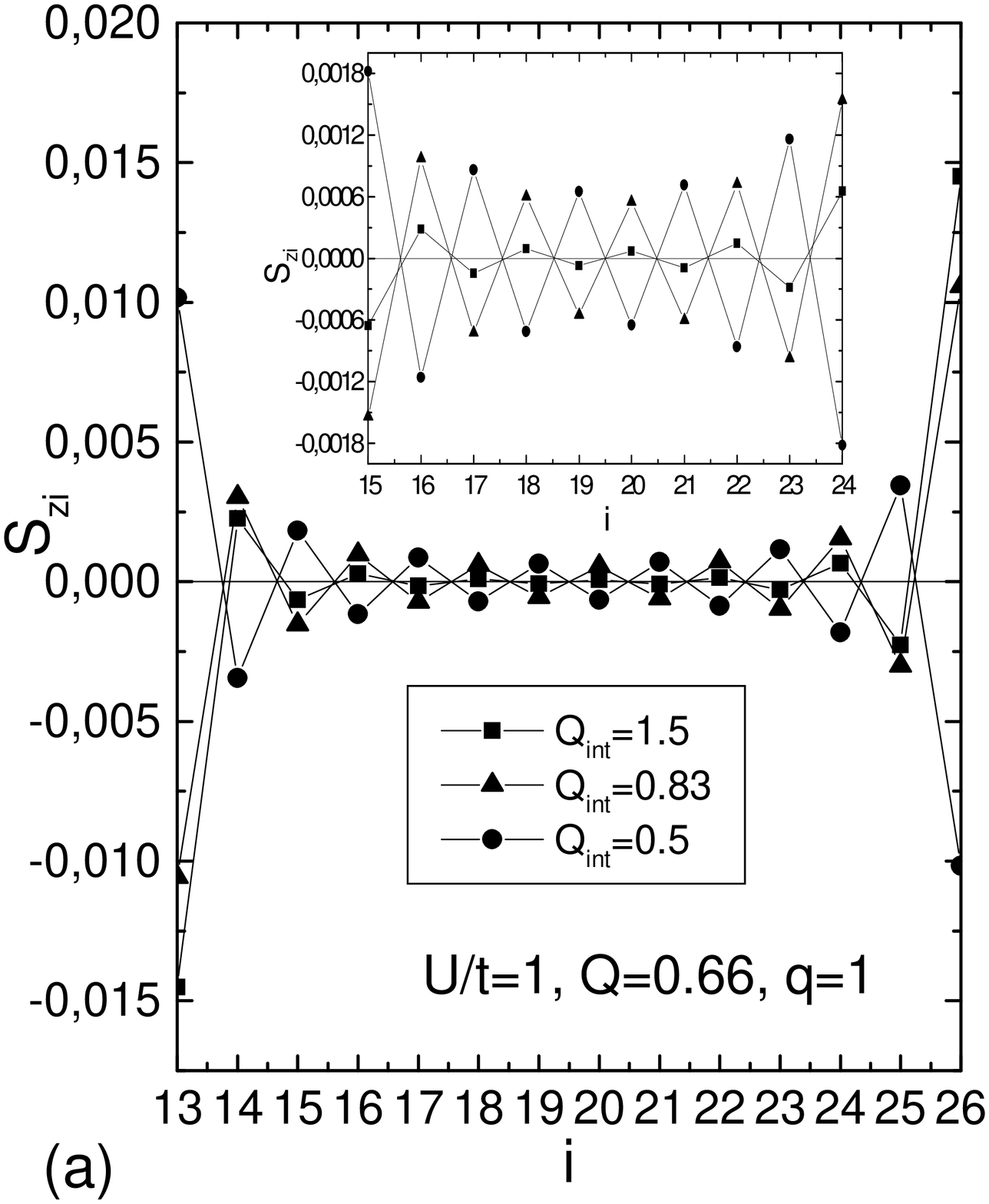} \includegraphics[width=7cm]{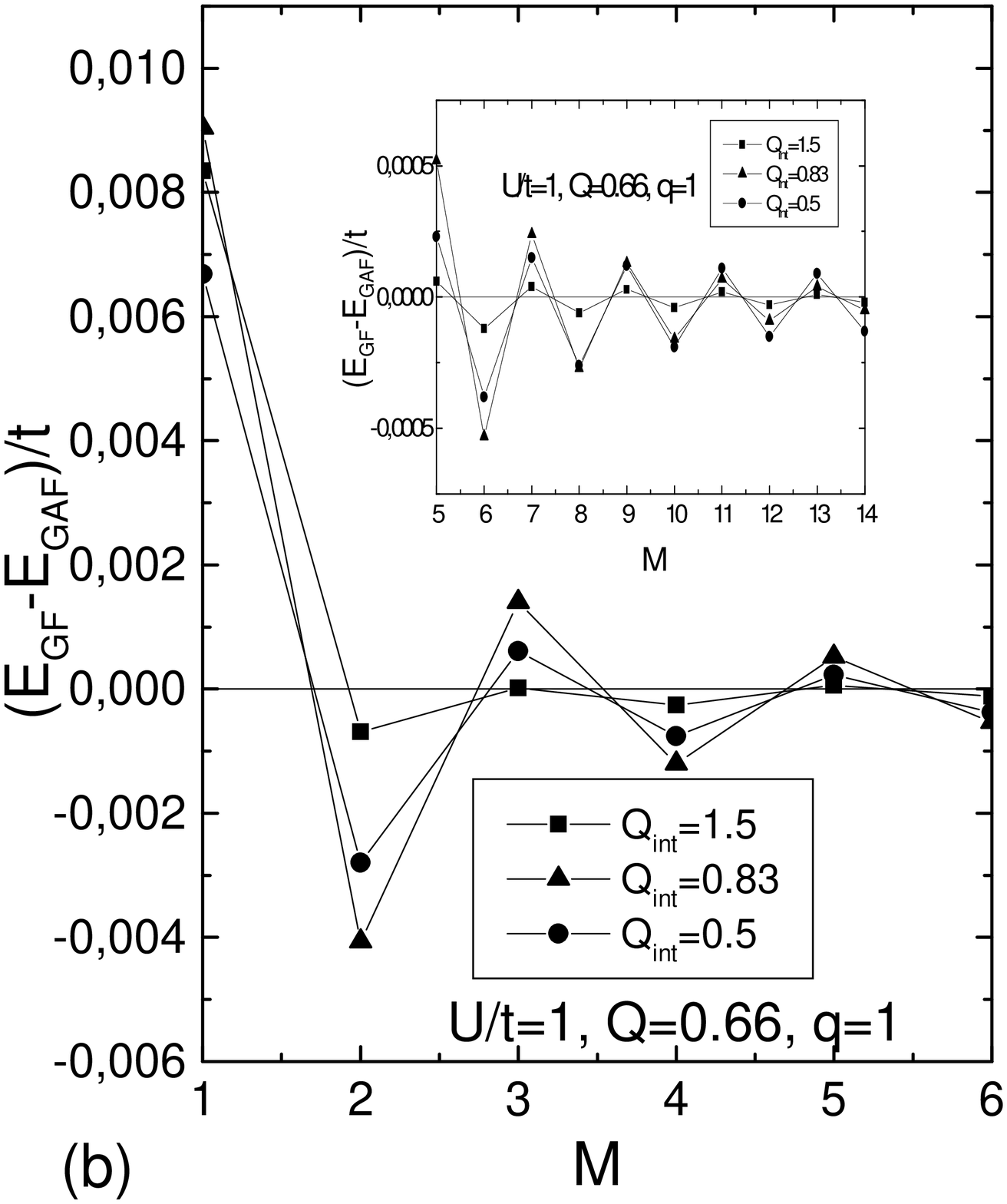}
\end{center}
\caption{(a) $S_{zi}=(n_{i\uparrow}-n_{i\downarrow})$ as a function of
  $i$ for $N$=12, $M$= 14, $U/t$=1, $Q$ =0.66, $q$=1, and three
  different values of $Q_{int}$ in the AF alignments. Inset: $S_{zi}$
  for the same parameters and $i$ from 15 to 24. (b) The energy
  difference (E$_{GF}$ - E$_{GAF}$)/$t$ as a function of the spacer
  planes number ($M$) for $N$=12, $U/t$=1, $Q$ =0.66, $q$=1, and
  $Q_{int}$ as in Fig 2a.in the AF alignments. Inset: (E$ _{GF}$ -
  E$_{GAF}$)/$t $ for the same parameters and $M$ from 5 to 14$.$}
\label{fig3}
\end{figure}

\begin{figure}[ht]
\begin{center}
\includegraphics[width=7cm]{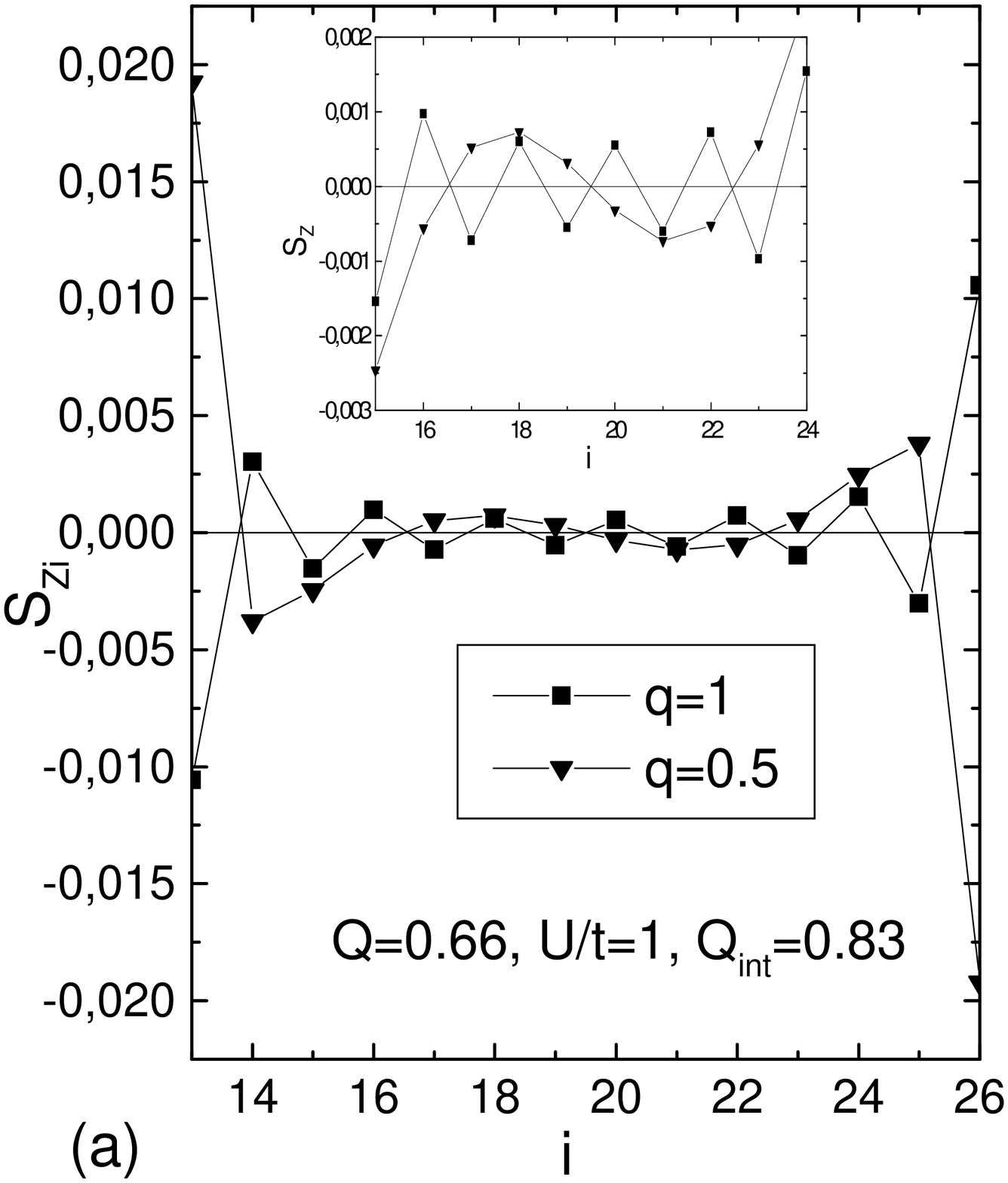}
 \includegraphics[width=7cm]{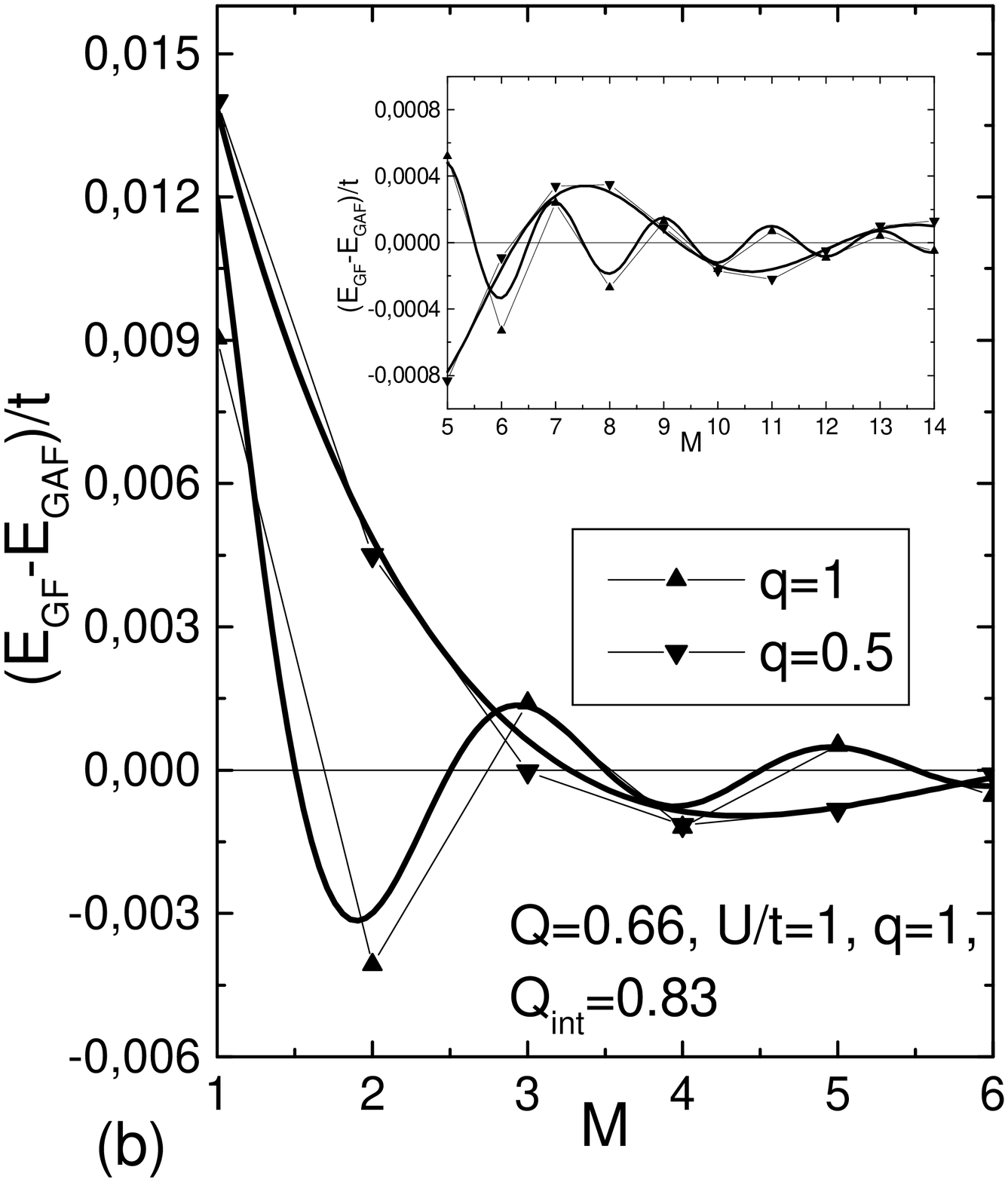}
\includegraphics[width=7cm,angle=-90]{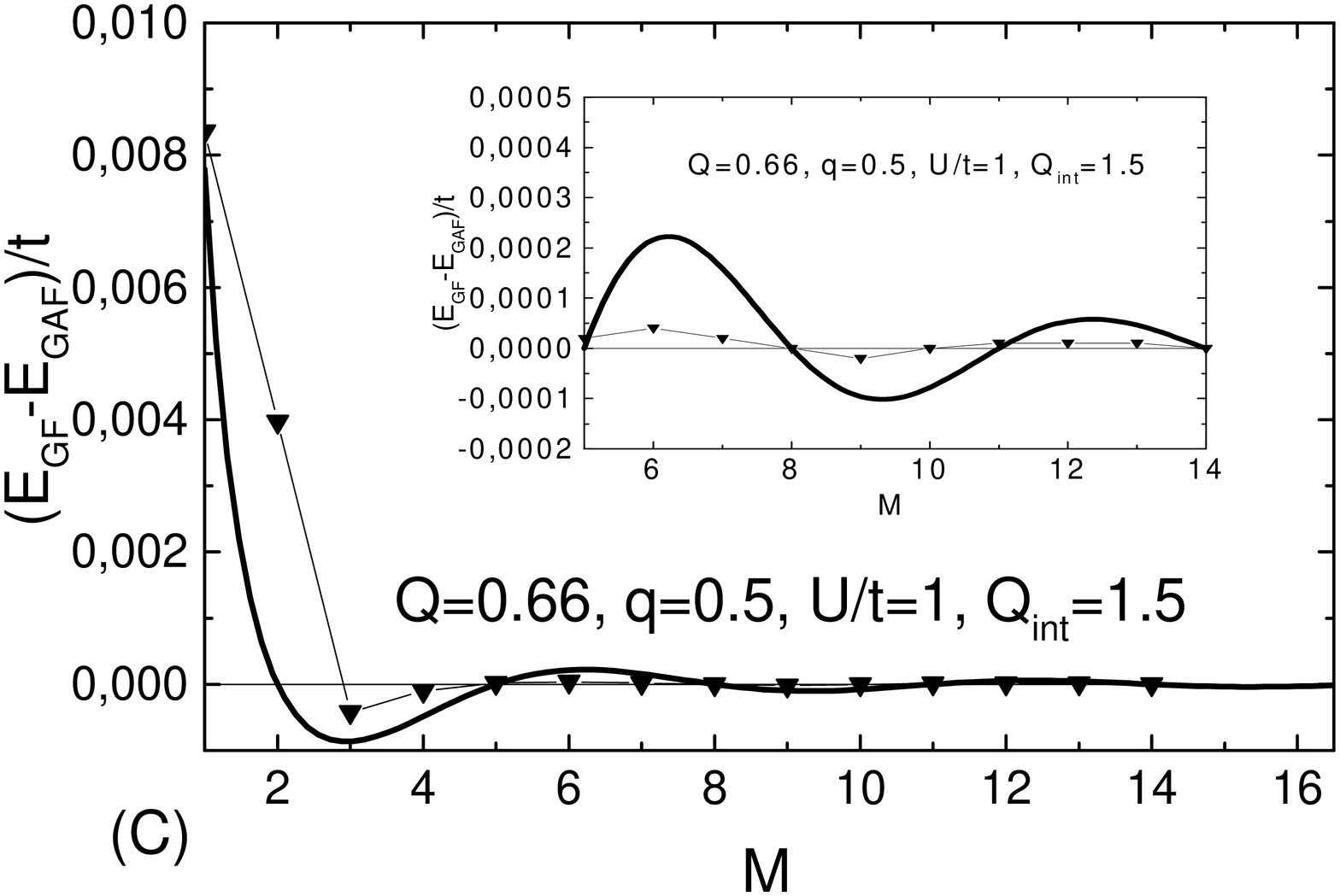}
\end{center}
\caption{(a) $S_{zi}=(n_{i\uparrow}-n_{i\downarrow})$ as a function of
  $i$ for $N$=12, $M$= 14, $U/t$=1, $Q$ =0.66, $Q_{int}$= 0.83, and
  two different values of $q$ in the AF alignments. Inset: $S_{zi}$
  for the same parameters and $i$ from 15 to 24. (b) The energy
  difference (E$_{GF}$ - E$_{GAF}$)/$t$ as a function of $M$ for
  $N$=12, $U/t$=1, $Q$ =0.66, $Q_{int}$= 0.83, and two different
  values of $q$. Inset: (E$_{GF}$ - E$_{GAF}$) for the same parameters
  and $M$ from 5 to 14. The solid lines correspond to the first term
  in the stationary phase approximation (SPA). (c)The energy
  difference (E $_{GF}$ - E$_{GAF}$)/$t$ as a function of $M$ for
  $N$=12, $U/t$=1, $Q$ =0.66, $Q_{int}$= 1.5, and $q$=0.5. Inset:
  (E$_{GF}$ - E$_{GAF}$) for the same parameters and $M$ from 5 to
  14. The solid lines correspond to the first term in the stationary
  phase approximation (SPA).}
\label{fig4}
\end{figure}

In this paper we focus on the CT effects on the separator (planes
$i=13$ to $ 26$). To study the magnetization in these planes we define
first $ S_{zi}=(n_{i\uparrow}-n_{i\downarrow})$ the magnetic
polarization at each PM plane. We show the effect of $Q_{int}$ on the
magnetization inside the PM displaying $S_{zi}$ as function of $i$ for
the antiferromagnetic arrangements (Fig.~\ref{fig3}(a)) of the FF
layers. We observe that the amplitude of the oscillations of\ $S_{zi}$
at the interface increases with $Q_{int}$ while the opposite occurs
far from the interface, a change of sign of $S_{zi}$ occurs for
different values of $ Q_{int}$ as one can observe in
Fig. \ref{fig3}(a). For F or AF alignments we calculate the ground
state energies E$_{GF}$ and E$_{GAF}$ respectively for each value of
$M$. The resulting exchange coupling $E_{x}=$ $E_{GF}$ - $E_{GAF}$ as
a function of $M$ is shown in Fig.~\ref{fig3}(b).  The period of
oscillations of\ $S_{zi}$ and $E_{x}$ can be identified with the
extremal spanning vectors of the Fermi surface of the spacer
($k_{zF}$).  This is shown in Fig.~\ref{fig4}(a) and (b), where one
can observe that changing $q$ and consequently the Fermi energy
changes the period of oscillations accordingly. \ In
Fig.~\ref{fig4}(b) the continuous line corresponds to the first term
in the stationary phase approximation (SPA) \cite{Ma}. The SPA first
term has the form $E_{x}=\frac{C}{(M+1)^{2}}\sin [2k_{zF}(M+1)+\varphi
]$, where $C$ and $\varphi $ are parameters that we choose to fit the
numerical results in Fig. \ref{fig4}(b) and (c). Note that far from
the interface the SPA reproduces quite well the numerical results (see
inset). In Fig. \ref{fig4}(c) we plot ($E_{GF}$ - $E_{GAF})/t$ as a
function of $M$ for $q=0.5$ and $Q_{int}=1.5,$ and compare with the
SPA approximation. We adjust $C$ and $\varphi $ so as to fit the first
point of the numerical result ($M=1$). It can be seen that the later
attenuate rapidly with $M$ much faster than the SPA result. This
strong attenuation has been found in
La$_{0.66}$Ba$_{0.34}$MnO$_{3}$-LaNiO$_{3}$-La$_{0.66}$Ba$
_{0.34}$MnO$_{3}$ trilayers by Nikolaev et al. in Ref. \cite{Ni} and
attributed to damping caused by strong electron scattering in the
non-magnetic layer.

\section{DISCUSSION AND CONCLUSION}

We have presented a simple four parameter model to describe the
coupling between two magnetic half metallic perovskites (as for
example La$_{1-x}$Sr$ _{x}$MnO$_{3}$.) separated by a metallic one
(LaNiO$_{3}$). The coupling is strongly affected by the charge
transfer at the interface due to the different effect of the Coulomb
interaction in the two materials. Recent two orbital calculations by
Ohsawa et al.\cite{ohsawa} using a potential difference to distinguish
between manganite and spacer do not show these effects. For example in
Fig.2 the broad line indicates the charge values at each atomic plane
obtained using simply a potential difference adjusted to reproduce the
values of $Q$ and $q$ (potential difference $\Delta /t=1.12)$.  We can
see that the result is a monotonous increase at the left interface
from 0.66 to 1 and the opposite at the right interface.

Of the four parameters, only $U/t$ is not determined by the materials
characteristics: $Q$ controls the number of $e_{g}$ electrons in the
manganites, $Q_{int}$ charge at the interface, is a parameter that
represents the possible differences between bulk and interface, $q$
($q=1$ in this calculation) controls the number of electrons in the
separator. The single free parameter $U$, is hard to estimate: it
contributes to the Coulomb interaction as the sum of point charge
interactions in the lattice.  This would be a good approximation if
the charge distribution were punctual or if there were no overlap
between charges. This is far from reality, an ab-initio of the
effective charges filling the atomic basin according to Bader's theory
gives significantly smaller values than the ideal ionic
value. Hybridization with the nearest oxygen ions affects strongly the
space charge distribution in non overlapping volumes. This reduces
significantly the values of $U.$ The ab-initio calculations were
performed using the full-potential linearized/augmented plane wave
plus local orbital (L/APW+lo) method, as implemented in the WIEN2K
code\cite{WIEN1,WIEN2,WIEN3}. The exchange-correlation effects were
treated within the GGA (generalized gradient approximation) using the
Perdew-Burke-Ernzerhof form \cite{WIEN4}.  For the calculation of
charges, we used the Bader's definition of atomic basins\cite{Bader}
calculated with the electronic densities obtained from the ab-initio
calculations. The atomic basins were calculated with the total
electronic density and then used to integrate the number of up and
down electrons assigned to each atom.

In table I we show the results for two cases: LaMnO$_{3}$ in the
A-type antiferromagnetic phase, and CaMnO$_{3}$ in a cubic G-type
antiferromagnetic phase. While in CaMnO$_{3}$ all oxygen atoms are
equivalent (O1 in table I), in LaMnO$_{3}$ there are two inequivalent
oxygen atoms: the atoms within each ferromagnetic plane (O1) and the
atoms located between two opposite ferromagnetic planes(O2).

\begin{table}[ht]
  \centering
  \begin{tabular}{|c|c|c|c|c|}
    \hline
    & \multicolumn{2}{l|}{LaMnO$_3$} & \multicolumn{2}{l|}{CaMnO$_3$} \\
    \hline
    & Charge & MM & Charge & MM \\
    \hline
    La/Ca & 2.08 & 0.00 & 1.66 & 0.00 \\
    \hline
    Mn & 1.66 & 3.53 & 1.81 & 2.92 \\
    \hline
    O1 & -1.25 & 0.11 & -1.15 & 0.00 \\
    \hline
    O2 & -1.24 & 0.00 & - & - \\
    \hline
  \end{tabular}
  \caption{Charges and magnetic moments (MM) obtained from the
    ab-initio calculations and using the Bader's definition of atomic
    basins\cite{Bader}.}
\end{table}

The results presented in table I indicate that the charge transfered by 
substitution of La by Ca is transfered not only to Mn as one purely ionic 
picture would indicate but it is transfered evenly to Mn and O. This is a consequence
 of the strong covalent bond tetween the transition metal and Oxigen.
This result agrees with the conclusions obtained by Raebiger et al. 
reported in NATURE\cite{rae} and extends their results to perovskites. 
Another factor that reduces $U$ is screening. We have also modified
the calculation including a term $e^{-\lambda \left\vert
    R_{i}-R_{j}\right\vert }$ in the interaction, and used different
values of $\lambda $ which is also difficult to estimate in the ionic
perovskites. We do not include here a study of the relaxation of the
interface plane which could enhance the charge effects at the
interface. We have also estimated the corrections due to different
hopping magnitude for the spacer and found that these corrections do
not produce qualitative changes in the results.

A quantitative calculation describing these type of interfaces is
still far from reach. It should include at least a two orbitals
description of the electronic structure as well as the effects of
strain and defects at the interface. We believe that our study of the
effect of coulomb interactions contributes to the understanding of
some qualitative aspects of their properties.

In conclusion, we have shown the charge transfer at the interface
between a half metallic and a metallic oxide is quite anomalous due to
the ionic nature of the materials under study. These anomalies
transfer to the different properties of the layered structures, as for
example conductance trough the layers, magnetization in the spacer and
magnetic coupling between the ferromagnetic layers. We have analyzed
here the magnetic exchange in a trilayer formed by two half metals
spaced by a metal.

\begin{center}
\textbf{Acknowledgments}
\end{center}
B. A., J. D. F., and R. A. are supported by the Consejo Nacional de
Investigaciones Cient\'{\i}ficas y T\'{e}cnicas (CONICET).

\end{document}